\documentstyle[epsfig, 12pt]{article}
\title{Low Energy Vortex Dynamics in Abelian Higgs Systems. }
\author{Wifredo Garc\'{\i}a Fuertes \\Departamento de F\'\i sica, Facultad de Ciencias, \\ Universidad de Oviedo \\ SPAIN\\and\\Juan Mateos Guilarte \\Departamento de F\'\i sica, Facultad de Ciencias, \\ Universidad de Salamanca \\ SPAIN}
\date{}
\newcommand{\beq}{\begin{equation}}
\newcommand{\eeq}{\end{equation}}
\newcommand{\bdm}{\begin{displaymath}}
\newcommand{\edm}{\end{displaymath}}
\begin{document}
\maketitle
\begin{abstract}

The low energy dynamics of the vortices of the Abelian Chern-Simons-Higgs system is investigated from the adiabatic approach. The difficulties involved in treating the field evolution as motion on the moduli space in this system are shown. Another two generalized Abelian Higgs systems are discusssed with respect to their vortex dynamics at the adiabatic limit. The method works well and we find bound states in the first model and scattering at right angles in the second system.
\end{abstract}
\section{Introduction.}
Since their discovery by Nielsen and Olesen \cite{bib:no}, the vortex solutions present in the Abelian Higgs Model have been used, beyond their original purpose as vehicles of the strong forces, in a variety of contexts. They have been found useful, for example, to describe cosmic strings; also, because the energy of static configurations in the AHM can be interpreted as the Ginzburg-Landau theory for superconducting materials, these topological solutions correspond to the magnetic flux tubes appearing in type II superconductors. The spectrum of potentially relevant vortices in condensed matter physics has recently be broadened by the discovery of a new class of outstanding cousins of the AHM solutions; the topological and non-topological solitons arising in several Chern-Simons-Higgs gauge systems. The need to include a Chern-Simons term in the treatment of three-dimensional gauge theories was first advocated by Jackiw and Templeton \cite{bib:jt} who were studying the radiative corrections to spinorial electrodynamics. The most remarkable effects of this term are the generation of a topological photon mass compatible with gauge invariance \cite{bib:jt},\cite{bib:jdt}, and the statistical transmutation of particles coupled to the gauge field \cite{bib:poly}. The Higgs mechanism in Maxwell-Chern-Simons electrodynamics was first investigated in \cite{bib:pr}, but although there are vortices in this system, they are not self-dual \cite{bib:dev}. The simplest way of achieving a self-dual limit is to renounce the Maxwell term and use the effective long-wavelength model introduced in \cite{bib:jlw},\cite{bib:hkp}. Self-duality with the Maxwell term is also possible, but then supplementary scalar fields become necessary \cite{bib:bz}. The CSH vortices could provide a theoretical model for describing physically distinguished objects such as Laughlin quasiparticles or quasiholes \cite{bib:pran} and the vortices of the still poorly understood high-$T_c$ superconductors \cite{bib:wil}. For this reason, any insight into their interactions and dynamical properties is of interest.

The non-linear nature of field equations having soliton-like solutions makes it almost impossible to study the dynamics of topological defects in full detail. A brilliant idea from Manton \cite{bib:man} in some cases allows an analytical approach to the problem: he showed that the low energy scattering of BPS monopoles can be traced back to geodesic motion in the moduli space of these self-dual solutions for fixed magnetic charge. The method has been generalized by Manton himself and others according to the following scheme: the adiabatic limit in the dynamics of topological defects is given by a Lagrangian system describing the motion of a particle in the moduli space of self-dual solutions ${\cal M}_n$. The mechanical kinetic energy comes from the terms which are quadratic in time-derivatives of the field theory action.
Linear terms in time-derivatives of the fields lead to a linear term in the velocity in the mechanical Lagrangian, inducing a Lorentz force.
Finally, the static part of the field theoretical energy produces the mechanical potential energy. With this
procedure, the adiabatic method fixes the geometric structure of the moduli space ${\cal M}_n$: the zeroth, first- and second- order terms in time-derivatives in the field theory action, respectively supply the definition of the manifold ${\cal M}_n$ itself, its complex structure´, and its metric. This way of proceeding has been successfully applied in a variety of models: there are, for example, some works on the AHM, both at the self-dual point \cite{bib:rub}, \cite{bib:sam} or away from it \cite{bib:stua}, that lead to second-order dynamics without a Lorentz term, but Manton has also shown \cite{bib:mant} how the same vortices can be embodied in a theory that has purely first-order dynamics.

The adiabatic method has also been applied to the analysis of the scattering of CSH vortices in \cite{bib:dzi}. In this system, the approach runs into difficulties and the reasons why the method fails are also pointed out in the same work. Regarding this problem, we realize that the same moduli space of self-dual vortices can be part of different field theories, as Manton discovered for the Nielsen-Olesen vortices at critical coupling. We shall therefore start from a fixed moduli space, ${\cal M}_n$, of topological vortices and search for ``simple" Lagrangians such that the points of ${\cal M}_n$ will be absolute minima of the field energy. Here we take the simplicity requirement as having the most natural dynamics, i.e. that associated with Manton`s approach. We shall see that the simultaneous existence of first- and second-order time-derivative terms in the theory, also present in CSH models, leads to a complex dynamical system on the moduli space and that difficulties appear in a complete analytical treatment. In this paper, however, we shall study two generalized Abelian Higgs models that share the same moduli space of vortices with the CSH system. The first of the models is non-relativistic and first-order vortex dynamics arises, captured at the adiabatic limit. The other model is relativistic and the vortices evolve according to second-order dynamics. Comparison with the application of the adiabatic method to the CSH system helps to clarify the origin of the problems found in this model. In our analysis 
we find a 
universal kind of behaviour: the low energy dynamics of topological CSH vortices in the non-relativistic model resembles the adiabatic limit of the Ginzburg-Landau theory proposed in \cite{bib:mant}. Topological vortices in the generalized AHM scatter at low energies as do Nielsen-Olesen vortices in the AHM. When first- and second-order dynamics are entangled, the adiabatic limit becomes very cumbersome.

The rest of the paper is organized as follows. In the next Section the CSH vortices are introduced, the issue of low energy dynamics is addressed and the difficulties made clear. This is based on previous work performed in \cite{bib:kim} and \cite{bib:dzi}. The next two sections are devoted to studying two alternative dynamics for the same vortices: new first- and second-order vortex dynamics, arising in a non-relativistic model and a relativistic one, are discussed. Some further comments and brief general conclusions are offered in the last section.
\section{The adiabatic limit and CSH vortex dynamics.}
\subsection{The moduli space of vortices in the CSH model.}
The action of the Abelian Chern-Simons-Higgs gauge system is \cite{bib:jlw},\cite{bib:hkp}
\beq
S=\int d^3x\{\frac{\kappa}{4}\varepsilon^{\alpha\beta\gamma}A_\alpha F_{\beta\gamma}+\frac{1}{2}D_\mu\phi^* D^\mu\phi-\frac{\lambda}{8}|\phi|^2(|\phi|^2-v^2)^2\}\label{eq:1}
\eeq
where the spacetime is three-dimensional, the metric is $g_{\mu\nu}={\rm diag}(1,-1,-1)$, and the covariant derivative is $D_\mu\phi=\partial_\mu\phi +ieA_\mu\phi$. The Lagrangian is quasi-invariant against the gauge transformations
\beq
\phi\rightarrow e^{ie\Lambda}\phi\;\;\;\;\;\;\;\; A_\mu\rightarrow A_\mu-\partial_\mu\Lambda\label{eq:2}
\eeq
Bearing in mind that
\begin{eqnarray}
\frac{\kappa}{4}\varepsilon^{\mu\nu\rho}A_\mu F_{\nu\rho}&=&\frac{\kappa}{2}\varepsilon_{kl}\dot{A}_kA_l+\kappa A_0 F_{12}+{\rm divergence}\nonumber\\ D_0\phi^*D^0\phi&=&(\partial_0|\phi|)^2+|\phi|^2(eA_0+\partial_0\arg (\phi))
\end{eqnarray}
and eliminating $A_0$ by means of the Gauss law coming from (\ref{eq:1})
\beq
A_0=-\frac{\kappa F_{12}}{|\phi |^2}-\frac{1}{e}\partial_0\arg(\phi)\label{eq:3}
\eeq
the action separates into kinetic and potential parts as follows
\beq
S=\int dt\{ T-V\}\label{eq:4a}
\eeq
\beq
T=\int d^2x\{\frac{1}{2}\dot{\varphi}^2+\frac{\kappa}{2}\varepsilon_{kl}\dot{A}_kA_l-\frac{\kappa}{2e}\dot{\Theta}F_{12}\}\label{eq:4b}
\eeq
\beq
V=\int d^2x\{\frac{1}{2}\frac{\kappa^2}{e\varphi^2}F^2_{12}+\frac{1}{2}D_k\phi^*D_k\phi +\frac{\lambda}{8}\varphi^2(\varphi^2-1)^2\}\label{eq:4c}
\eeq
where $\phi=\varphi e^{\frac{i}{2}\Theta}$. Passing to the Hamiltonian formalism, we find
\beq
H=\int dt\{K+V\}
\eeq
with
\beq
K=\frac{1}{2}\int d^2x\dot{\varphi}^2.
\eeq

We shall first focus on static configurations. For these, $L=-V$, $H=V$ and the finiteness of the energy requires $\phi(\vec{x})\rightarrow 0\;{\rm or}\; v$ when $|\vec{x}|\rightarrow \infty$. In this paper, we limit ourselves to the second case, i.e. we shall work on the configuration space
\beq
{\cal C}=\{\Gamma\equiv(\phi, A_k)/\dot{\Gamma}=0, E[\Gamma]<\infty, \phi|_{|\vec{x}|\rightarrow\infty}=v\}
\eeq
\indent Each configuration in ${\cal C}$ gives rise to a map $\phi_{\infty} :S^1_{\infty}\rightarrow U(1)$ and is therefore associated with an integer, the winding number $n$ of $\phi_\infty$. As a consequence, ${\cal C}=\cup_{n\in{\bf Z}}{\cal C}_n$ and a topological superselection rule arises: time evolution cannot change the initial winding number. Furthermore, because $D_k\phi$ must vanish at infinity, the magnetic flux of the configurations in ${\cal C}_n$ is given by ${\displaystyle \Phi_M\equiv -\int d^2x F_{12}=\frac{2\pi n}{e}}$.

Our interest lies in the solutions belonging to ${\cal C}_n$, which are topological $n$-vortices. Although the theory also includes another class of very interesting non-topological solutions with a vanishing asymptotic scalar field, there is evidence that such non-topological solutions can be understood as assemblies of vortices mixed with some basic non-topological defects \cite{bib:smrk}. Hence, the dynamics of this kind of solution only differs from that of the topological vortices in the effect of the vortex-defect interaction, an issue to be dealt with elsewhere. In order to render $V$ extremal in ${\cal C}$, the Bogomolnyi trick is useful
\begin{eqnarray}
V&=&\int d^2x\{\frac{1}{2}[\frac{\kappa F_{12}}{e\varphi}\mp\frac{e^2}{2\kappa}\varphi (\varphi^2 -v^2)]^2+\frac{1}{2}|D_1\phi\pm iD_2\phi|^2+\nonumber\\ &+&\frac{1}{8}(\lambda-\frac{e^4}{\kappa^2})\varphi^2 (\varphi^2 -v^2)^2\}\pm\frac{ev^2}{2}\Phi_M\label{eq:7}.
\end{eqnarray}
\indent There is a critical point at $\lambda=\frac{e^4}{\kappa^2}$ where the contribution of the third term vanishes and a global lower bound to the energy arises: $V\geq\pi v^2|n|$ for any configuration in ${\cal C}_n$. The bound is saturated if and only if the first order equations
\begin{eqnarray}
eF_{12}&=&\pm\frac{m^2}{2}\frac{\varphi^2}{v^2}(\frac{\varphi^2}{v^2}-1)\label{eq:8a}\\D_1\phi&\pm &iD_2\phi=0  \label{eq:8b},
\end{eqnarray}
where $m=\frac{e^2v^2}{\kappa}$, are satisfied; solutions of (\ref{eq:8a}), (\ref{eq:8b}) are also solutions of the Euler-Lagrange equations. We see that by replacing the Maxwell term by the Chern-Simons term, self-duality requires a potential of sixth order in the modulus of the Higgs field. Below we fix the upper sign in these equations and work on ${\cal C}_n$ with $n>0$; the opposite choice would lead to analogous antivortices with $n<0$.

Using the Poincar\'e $\bar{\partial}$ lemma it is possible to prove that the Higgs field of the non-singular solutions of (\ref{eq:8b}) has exactly $n$ zeroes and that away from them the phase $\Theta$ is regular \cite{bib:w},\cite{bib:jft}. Furthermore, near a zero $\vec{q}$ of order $r$, the field behaviour is
\beq
\varphi\simeq c|\vec{x}-\vec{q}|^r\;\;\;\;\;\;\;\; \Theta\simeq 2r\theta(\vec{x}-\vec{q})\label{eq:14a}
\eeq
 $\theta(\vec{x})$ being the polar angle of $\vec{x}$. The self-duality equations over ${\bf R}^2-\{\vec{q}_1, \vec{q}_2,\ldots, \vec{q}_n\}$ are
\begin{eqnarray}
\nabla^2u&=&m^2 e^u(e^u-1)\label{eq:9a}\\eA_k&=&-\frac{1}{2}(\partial_k\Theta+\varepsilon_{kj}\partial_ju)\label{eq:9b}
\end{eqnarray}
with $u=\ln (\frac{\varphi}{v})^2$. Observe that with respect to the corresponding equations in the AHM there is 
an additional factor, $e^u$, on the right hand side of equation (15).  

The manifold of solutions of (\ref{eq:8a})-(\ref{eq:8b}) on ${\cal C}_n$  modulo the group of gauge diffeomorphisms is the $n$-vortex moduli space ${\cal M}_n$. As proved by Wang \cite{bib:w}, ${\cal M}_n$ is the smooth manifold of unordered $n$-points in ${\bf C}$: ${\cal M}_n=\frac{{\bf C}^n}{\Sigma^n}$, where $\Sigma^n$ is the symmetric group of $n!$ elements. This is so because the $n$ zeroes of $\phi$ in ${\bf C}\simeq{\bf R}^2$ determine a unique solution, up to permutation and gauge equivalence. A system of ``good'' coordinates in ${\cal M}_n$ is provided by the coefficients of the complex monic polynomial of degree $n$ whose roots are the zeroes of $\phi$: $P(z)=z+a_1z^{n-1}+...+a_n$, with $P(z_a)=\phi(z_a)=0$ for $z_a=q_a^1+iq_a^2$, $a=1,2,...,n$. Had we chosen the centres of the vortices $z_a$ as a system of coordinates in ${\cal M}_n$, singularities would have appeared when two zeroes coincided. From a physical viewpoint, the structure of ${\cal M}_n$  shows that at the self-dual limit the scalar attractive force and the gauge repulsive force compensate each other mutually and hence the static self-dual solutions consist of systems of non-interacting vortices.
\subsection{The dynamics of slowly moving vortices.}
We now address the issue of the time evolution of a self-dual system of vorticity $n$. Because the time-dependent field equations are too difficult to solve, it is necessary to restrict the problem in such a way that an approximate treatment is feasible. The most natural restriction is to limit ourselves to the case of very slowly evolving fields so that we can address the problem with Manton's adiabatic method: the point is that the solutions $\Gamma[\vec{x},t]$ with $\dot{\Gamma}$ small essentially describe the motion of the individual vortices. We can thus identify $\Gamma[\vec{x},t]$  with a curve $\{\vec{q}_a(t)\}$ in ${\cal M}_n$,  i.e.
\beq
\Gamma(\vec{x};t)=\Gamma(\vec{x};\vec{q}_a(t))\;\;\;\;\;\;\; \dot{\Gamma}(\vec{x};t)=\frac{\partial\Gamma(\vec{x};\vec{q}_a(t))}{\partial q_a^k}\dot{q}_a^k\label{eq:10}.
\eeq
\indent The field-theoretical problem is transmuted to a $2n$-dimensional mechanical one: introduction of (\ref{eq:10}) into (\ref{eq:4a}) and integration to the whole plane afford a Lagrangian $L=T(\vec{q}_a,\dot{\vec{q}}_a)-V(\vec{q}_a)$ whose variational equations admit as a solution the curve $\{\vec{q}_a(t)\}$ in ${\cal M}_n$ corresponding to some given initial conditions. In fact, on the moduli space, $V(\vec{q}_a)=\pi n v^2$ and the only important term of $L$ is the kinetic one.

To carry out this program, the first step is to unequivocally fix the form of $\Gamma[\vec{x};\vec{q}_a]$ i.e, to fix the gauge by defining $\Theta(\vec{x};\vec{q}_a)$ locally on the moduli. This gauge fixing must be done in such a way that the kinetic energy will be invariant not only against the group ${\cal G}$ of gauge diffeomorphisms but also against the enlarged group $\tilde{{\cal G}}$ of moduli-dependent gauge transformations: the dynamics cannot vary if we choose different gauges in different points of the moduli. However, despite this strong requirement, in the CSH model there is no restriction to our freedom to choose $\Theta(\vec{x};\vec{q}_a)$: the only requisite is to respect the boundary conditions in $\{\vec{x}=\vec{q}_a\}$ and $S^1_\infty$. The reason is that (\ref{eq:4b}) is invariant against moduli-dependent gauge transformations because we obtained that expresion from (\ref{eq:1}) merely by imposing the gauge-independent Gauss law (\ref{eq:3}). Hence, we can fix the gauge in the simplest form compatible with the boundary conditions
\beq
\Theta(\vec{x};\vec{q}_a)=2\sum_{a=1}^n\theta(\vec{x}-\vec{q}_a)\label{eq:11}
\eeq
i.e, by extending the known behaviour near the centres of the vortices to the whole plane. Introduction of (\ref{eq:11}) into the equation (\ref{eq:9b}) gives
\beq
eA_k(\vec{x};\vec{q}_a)=\varepsilon_{kj}\partial_j\xi(\vec{x};\vec{q}_a)\label{eq:12a}
\eeq
where
\beq
\xi(\vec{x};\vec{q}_a)=-\frac{1}{2}u(\vec{x};\vec{q}_a)+\sum_{a=1}^n\ln|\vec{x}-\vec{q}_a|   ;\label{eq:12b}
\eeq
Therefore, $\xi$  is regular on the whole ${\bf R}^2$, see (\ref{eq:14a}). Using (\ref{eq:12a}) and computing its time-derivative, we obtain 
\beq
\varepsilon_{kl}\dot{A}_kA_l=\partial_j(\dot{\xi}A_j)-\dot{\xi}(\partial_jA_j)
\eeq
Because the vector field of the vortices is transverse, the second term in (\ref{eq:4b}) is a global divergence and can be dropped. The third term in the kinetic energy can be written in the form, 
\beq
\frac{1}{2}\dot{\Theta}F_{12}=\varepsilon_{ij}\varepsilon_{kl}\partial_kA_l\sum_{b=1}^n\dot{q}_b^i\partial_j\ln |\vec{x}-\vec{q}_b|
\eeq
 using (\ref{eq:11}).
This expression is regular across the entire plane because $F_{12}$ vanishes at the center of the vortices. Proceeding by partial differentials, one can see that, besides an irrelevant divergence
\beq
\frac{1}{2}\dot{\Theta}F_{12}=2\pi\sum_{b=1}^nA_i(\vec{x})\dot{q}_b^i\delta(\vec{x}-\vec{q}_b)\label{eq:13}
\eeq
Following \cite{bib:sam}, we expand the modulus of the Higgs field near the $b^{th}$ vortex in the form, 
\beq
\frac{1}{2}u(\vec{x};\vec{q}_a)|_{\vec{x}\simeq\vec{q}_b}=\ln|\vec{x}-\vec{q}_b|+a_b+\vec{b}_b\cdot(\vec{x}-\vec{q}_b)+\ldots     ;\label{eq:13p}
\eeq
$a_b,\vec{b}_b$ are functions of the $\vec{q}^{\,\prime s}$, and then the value of the vector field at the center of that vortex
is
\beq
eA_k(\vec{q}_b;\vec{q}_a)=\varepsilon_{kj}[\sum_{a\neq b}\frac{q_b^j-q_a^j}{|\vec{q}_b-\vec{q}_a|^2}-b_b^j]\label{eq:14}
\eeq
for any solution of the vortex equations.

Substitution of (\ref{eq:14}) into formula (\ref{eq:13}) produces a term in the kinetic energy (\ref{eq:4b}) that involves only the $\vec{q}^{\,\prime s}$ and their time-derivatives. Unfortunately, it is not possible to obtain an explicit expression for the quadratic term in closed form, because integration to the whole plane requires detailed knowledge of $\varphi(\vec{x};\vec{q}_a)$ as a function of $\vec{x}$ and not only in the vicinity of each vortex. Because the exact solution of the system (\ref{eq:8a})-(\ref{eq:8b}) is unknown, the best thing that we can do is to write
the mechanical Lagrangian in the form
\beq
L=\frac{1}{2}\sum_{a,b=1}^n g_{ij}^{ab}\dot{q}_a^i\dot{q}_b^j-\frac{2\pi\kappa}{e}\sum_{b=1}^n\dot{q}_b^kA_k(\vec{q}_b;\vec{q}_a)-\pi v^2n\label{eq: aug}
\eeq
with $A_k$ given by (\ref{eq:14}) and
\beq
g_{ij}^{ab}=\int d^2x\frac{\partial\varphi}{\partial q^i_a}\frac{\partial\varphi}{\partial q^j_b}.\label{eq:16}
\eeq

The only possibility for integrating (\ref{eq:16}), giving a analytic expression for the metric,  is to consider the asymptotic regimes of either very close or very separated vortices. We now analyze the second case, in which the scalar field around each vortex is approximately radially symmetric, i.e. the $\vec{b}^{\prime s}$ are vanishingly small. This 
behaviour and the great distance among vortices guarantee that the vector field at their centres is negligible, see (\ref{eq:14}), and that the dynamics is governed by the quadratic term in $T$. Because $\varphi$  tends to $v$ exponentially when $|\vec{x}-\vec{q}|$ goes to infinity, it makes sense to write
\beq
\varphi (\vec{x})=\left\{\begin{array}{ccc}\varphi_1 (|\vec{y}_a|)&{\rm if}&|\vec{y}_a|<R_v\\v&{\rm if}&|\vec{y}_a|>R_v\end{array}\right.
\eeq
with $\vec{y}_a=\vec{x}-\vec{q}_a$ and $\varphi_1$ the magnitude of the Higgs field of the radially symmetric 1-vortex and $R_v$ its characteristic radius i.e., the radius of the circle in which $\varphi_1$ differs appreciably from $v$. It is easy then to see that 
\beq
g_{ij}^{ab}=\delta^{ab}\int d^2y\frac{y^iy^j}{r^2}(\frac{d\varphi_1}{dr})^2
\eeq
with $r=|\vec{y}|$, or
\beq
g_{ij}^{ab}=\delta^{ab}\delta_{ij}M,\;\;\;\;\;\;\;\;\; M=\frac{1}{2}\int d^2y(\frac{d\varphi_1}{dr})^2.\label{eq:17}
\eeq
Plugging the radial form of equation (\ref{eq:8b}) into this expression, 
\beq
\frac{d\varphi_1}{dr}=\frac{1+eA_\theta}{r}\varphi_1   ,
\eeq
we find
\beq
M=-\frac{e}{4}\int d^2x\varphi_1^2F_{12}
\eeq
But $\varphi_1^2<v^2$ and we conclude that $M<\frac{\pi v^2}{2}$. This is an inconsistent answer, implying that the inertia of each vortex is less than half its mass, which for the case $n=1$ leads to a conflict with relativistic invariance. Such nonsense strongly suggests that the adiabatic approach fails in the CSH model and needs to be improved. The critical analysis of the adiabatic method in the current model carried out by Dziarmaga \cite{bib:dzi} reveals the reason for the failure. We review this analysis in the next subsection.
\subsection{The improved adiabatic limit.}
Consider a general field theory with a field multiplet $(\psi_a)$  and a Lagrangian
\beq
{\cal L}=G_{ab}[\psi ]\dot{\psi}_a\dot{\psi}_b+K_a[\psi ]\dot{\psi}_a-{\cal H}[\psi ]\label{eq:18}
\eeq
where there are no time-derivatives inside the brackets. Assume that the static solutions of the field equations form a moduli space ${\cal M}$. $V=\int d^nx{\cal H}[\psi]$ takes the same constant value on each point of ${\cal M}$. Let $\{\lambda_A\}$ be a local system of coordinates in ${\cal M}$ and let $\varphi_a(\vec{x}; \lambda_A)$ denote the fields corresponding to the solution $\{\lambda_A\}$. At Manton's adiabatic limit, slow time evolution merely amounts to motion in the moduli space. Thus, time-dependence is exclusively due to variations in the $\{\lambda_A\}$ coordinates as functions of time:
\beq
\dot{\psi}_a=\frac{\partial\varphi_a}{\partial\lambda_A}\dot{\lambda}_A   .\label{eq:19}
\eeq
and hence, the effective Lagrangian
\beq
{\cal L}^{Manton}_{eff}=G_{ab}[\varphi ]\frac{\partial\varphi_a}{\partial\lambda_A}\frac{\partial\varphi_b}{\partial\lambda_B}\dot{\lambda}_A\dot{\lambda}_B+K_a[\varphi ]\frac{\partial\varphi_a}{\partial\lambda_A}\dot{\lambda}_A-{\cal H}[\varphi ].\label{eq:20}
\eeq
is obtained.

However, the true solutions of the time-dependent Euler-Lagrange equations are some configurations $\psi_a(\vec{x}, t)\neq\varphi_a(\vec{x}; \lambda_A(t))$. In principle, one could improve the adiabatic approach, even without knowledge of the exact solutions of the time-dependent non-linear field equations, by the inclusion of a linear term in $\dot{\lambda}_A$
\beq
\psi_a(\vec{x},t)=\varphi_a(\vec{x};\lambda_A(t))+\phi_a^B(\vec{x};\lambda_A(t))\dot{\lambda}_B(t)    .\label{eq:21}
\eeq
that accounts for the deformation of the static fields as a result of the motion. (\ref{eq:19}) now turns out to be 
\beq
\dot{\psi}_a(\vec{x}, t)=\frac{\partial\varphi_a}{\partial\lambda_A}\dot{\lambda}_A+\phi_a^A\ddot{\lambda}_A+\frac{\partial\phi_a^B}{\partial\lambda_A}\dot{\lambda}_A\dot{\lambda}_B.\label{eq:22}
\eeq
and the introduction of (\ref{eq:21}), (\ref{eq:22}) into (\ref{eq:18}) gives a very complicated expression:
\beq
{\cal L}_{eff}={\cal L}_{eff}^{(2)}+{\cal L}_{eff}^{(1)}\label{eq:23}
\eeq
where
\beq
{\cal L}_{eff}^{(1)}=K_a[\varphi ]\dot{\varphi}_a\label{eq:23a}
\eeq
\begin{eqnarray}
{\cal L}_{eff}^{(2)}&=&G_{ab}[\varphi ](\dot{\varphi}_a\dot{\varphi}_b+2\dot{\varphi}_a\dot{\Delta}_b+\dot{\Delta}_a\dot{\Delta}_b)+\nonumber\\ &+&\frac{\delta K_a}{\delta\psi_b}(\dot{\varphi}_a\Delta_b-\dot{\varphi}_b\Delta_a-\dot{\Delta}_b\Delta_a)-\frac{1}{2}\frac{\delta^2{\cal H}}{\delta\psi_a\delta\psi_b}\Delta_a\Delta_b\label{eq:23b}
\end{eqnarray}
with
\beq
\Delta_a=\phi_a^B\dot{\lambda}_B,\;\;\;\;\;\;\; \dot{\Delta}_a=\phi_a^A\ddot{\lambda}_A+\frac{\partial\phi_a^B}{\partial\lambda_A}\dot{\lambda}_A\dot{\lambda}_B.
\eeq
The question is whether or not this modification, which we shall call the improved adiabatic limit, has any physical meaning. There are three different cases:
\begin{description}
\item [A.] Assume that both $G_{ab}$ and $K_a$ are different from zero. Because $\dot{\varphi}_a$ is 
linear in $\dot{\lambda}_A$, all the terms in ${\cal L}_{eff}^{(2)}$ and ${\cal L}_{eff}^{(1)}$ are at least quadratic and linear in velocities, respectively. Integration over the whole plane of ${\cal L}_{eff}={\cal L}_{eff}^{(2)}+ {\cal L}_{eff}^{(1)}$ leads to a reduced mechanical Lagrangian taking the form:
\beq
L_{eff}=g_{AB}(\lambda)\dot{\lambda}_A\dot{\lambda}_B+h_A(\lambda)\dot{\lambda}_A
\eeq
which describes the motion on the moduli space ${\cal M}$. Because the energy is constant on
${\cal M}$ the static forces between vortices are null, i.e., $\ddot{\lambda}_A=\omega_{AB}(\lambda)\dot{\lambda}_B+o(\dot{\lambda}^2)$; the acceleration
is zero if the velocity is zero. We see that due to the linear term in $L_{eff}$, coming from ${\cal L}_{eff}^{(1)}$, $\omega_{AB}(\lambda)$ is not zero and hence $\Delta_a$ and $\dot{\Delta}_a$ are of the same order. 
Therefore, when $\omega_{AB}(\lambda)\neq 0$ one should replace the quadratic term in velocities in ${\cal L}_{eff}^{Manton}$ by ${\cal L}_{eff}^{(2)}$ and hence one should consider the improved
adiabatic limit. This is exactly the case in the CSH system:
\beq
G_{ab}\dot{\psi}_a\dot{\psi}_b=\frac{1}{2}\dot{\varphi}^2,\;\;\;\;\;\;\; K_a[\psi]\dot{\psi}_a=\frac{\kappa}{2}\varepsilon_{kl}\dot{A}_kA_l-\frac{\kappa}{2e}\dot{\Theta}F_{12}
\eeq
The dynamics of the CSH vortices at the improved adiabatic limit is governed by the mechanical Lagrangian $L_{eff}$, where $g_{AB}$ and $h_A$ are derived from ${\cal L}_{eff}={\cal L}_{eff}^{(2)}+ {\cal L}_{eff}^{(1)}$. This is a very difficult problem: first, it is not possible to give
a closed expression for the metric $g_{AB}(\lambda)$ because it depends not only on the vortex motion
in the moduli space, but also includes effects coming from the field deformations whose specification is beyond self-duality and makes the use of the Euler-Lagrange equations unavoidable. Moreover, there are Lorentz forces due to $h_A(\lambda)$ that would strongly disturb the possible geodesic motion in the metric $g_{AB}$.
\item [B.] Let us next consider the case where $G_{ab}$ is not zero but $K_a=0$. The mechanical Lagrangian is now
\beq
L_{eff}=g_{AB}(\lambda)\dot{\lambda}_A\dot{\lambda}_B
\eeq
and  $\omega_{AB}(\lambda)=0$. Then, $\ddot{\lambda}_A=-\Gamma_{ABC}(\lambda)\dot{\lambda}_B\dot{\lambda}_C$ and ${\cal L}_{eff}^{(2)}={\cal L}_{eff}^{Manton}$, up to quadratic order in the velocities. The remaining terms in ${\cal L}_{eff}^{(2)}$ are at least cubic in 
$\dot{\lambda}_A$. This statement is obvious for the terms in the first line of formula (\ref{eq:23b}) but one now needs to use the field equations derived from (\ref{eq:18}) to check that $\frac{\delta^2{\cal H}}{\delta\psi_a\delta\psi_b}$ is indeed proportional to $\dot{\lambda}_A$ at slow velocities. Thus, the modification  induced by formula (\ref{eq:21}) is negligible and the adiabatic limit is tantamount to geodesic motion in the moduli space. The Abelian Higgs Model obeys
 this situation with
\beq
G_{ab}\dot{\psi}_a\dot{\psi}_b=\frac{1}{2}(\dot{\phi}^*\dot{\phi}+\dot{A}_i\dot{A}_i)
\eeq
and the low energy dynamics of vortices becomes a palatable mechanical problem. In 
section 4 we shall discuss a generalized Abelian Higgs Model that also belongs to this type.
\item [C.] Finally, let us  consider the opposite situation: $G_{ab}=0$ but $K_a$ is not null. The key point is that in this case ${\cal L}_{eff}^{Manton}$ is linear in velocities and we do not need to consider the corrections induced in ${\cal L}_{eff}^{(2)}$ by deformations of the fields because they are at least of second order in $\dot{\lambda}_A$. The low energy dynamics is again captured by the adiabatic limit 
which now consists of a mechanical problem on the configuration space ${\cal M}$ with Lagrangian:
\beq
L_{eff}=h_A[\lambda]\dot{\lambda}_A    ,
\eeq
causing motion on the moduli space exclusively due to Lorentz forces.
The non-relativistic Ginzburg-Landau system analyzed in \cite{bib:mant} belongs to this type:
\beq
K_a[\psi]\dot{\psi}_a= i(\phi^* \dot{\phi}-\phi\dot{\phi}^*)+\frac{\kappa}{2}\varepsilon_{kl}\dot{A}_kA_l-\frac{\kappa}{2e}\dot{\Theta}F_{12}  .
\eeq
A generalization of this model in the same class will be studied in the next Section.
\end{description}

The conclusion is that the usual adiabatic approach is suitable for studying slow motion dynamics when the system is purely linear or quadratic in the time-derivatives of the fields, but not when there are terms of both types simultaneously. In this case, the approach  needs to be refined and this leads to a exceedingly complicated problem that does not admit any analytical treatment \cite{bib:dzi}.
\begin{displaymath}
*\ \ \ \ \ *\ \ \ \ \ *
\end{displaymath}

To close this Section we briefly discuss the issue of vortex CSH statistics, a rather paradoxical subject \cite{bib:kim}. For a topological CSH vortex at rest we have:
\beq
\Phi_M=\frac{2\pi}{e},\ \ \ \ \ \ Q=-\frac{2\pi\kappa}{e}, \ \ \ \ \ \ J=-\frac{\pi\kappa}{e^2}  .
\eeq
If we trust the standard computation of the statistical angle of two-dimensional anyons through the Aharonov-Bohm effect, we find that the CSH vortices correspond to a statistics $\nu=\frac{2\pi\kappa}{e^2}$ and the spin-statistics relation is $\nu=-2s$.; there is a minus sign with respect to the expected outcome. Nevertheless, in the adiabatic mechanical Lagrangian (\ref{eq: aug}), the term 
\beq
L_{stat}=-\frac{2\pi\kappa}{e^2}\varepsilon_{kj}\sum_b \dot{q}_b^k\sum_{a<b}\frac{q_b^j-q_a^j}{|\vec{q}_a-\vec{q}_b|^2}
\eeq
\beq
 =+\frac{2\pi\kappa}{e^2}\sum_{a<b}\frac{d}{dt}arg(\vec{q}_a-\vec{q}_b)
\eeq
should be interpreted as providing anyonic statistics, see \cite{bib:lerd}, for a statistical angle $\nu=-\frac{2\pi\kappa}{e^2}=2s$. We find the right answer at the adiabatic limit, whereas application of the AB method to extended distributions of electric and magnetic charge fails.
\section{The non-relativistic linear model.}
The paradigm of linear gauge theory in the time-derivatives is the non-relativistic model of Jackiw and Pi \cite{bib:jp}, which describes the minimal coupling between the non-linear Schr\"{o}dinger matter field and the Chern-Simons gauge field
in $(2+1)$-dimensions. Although this model contains self-dual vortices, these are quite different from that considered in the previous section. In the Jackiw-Pi theory one only has the symmetric phase, constructed on the unique vacuum $\phi=0$, and the vortices are non-topological even if the magnetic flux is integer. In fact, the JP model is the non-relativistic limit of the CSH system and the JP vortices are the corresponding limit of the non-topological CSH vortices; the topological vortices disappear from the spectrum in the non-relativistic regime and the flux quantization is due to the inversion properties of the Liouville equation rather than to topological reasons (the JP model enjoys conformal invariance and the vortex equations become equivalent to the Liouville equation). As we shall see, to have true self-dual topological vortices, the original JP theory must be modified with due care.
\subsection{The generalized Ginzburg-Landau theory.}
We shall now discuss a non-relativistic model with both symmetric and asymmetric phases. The new system is a generalization of the model analyzed by Manton in Reference \cite{bib:man}; the crucial difference is that the moduli space of topological vortices is now the same as in the CSH theory instead of being the moduli space of Ginzburg-Landau vortices. Of course, we shall find first-order vortex dynamics rather than the awkward situation of the CSH model. Before, however, we must deal with the tricky question of making non-relativistic dynamics compatible with the spontaneous symmetry breakdown of $U(1)$ invariance.
Even though it is possible to build a non-symmetric vacuum in the JP theory, the fields cannot reach it asymptotically because that would lead to pathologies; namely, infinite charges and a misdefinition of the canonical formalism. Barashenkov and Harin \cite{bib:bh} traced the origin of the problem back to the underlying pure scalar model in 1+1 dimensions and found that a possible loophole is to multiply the $\dot{\phi}$ term of the Lagrangian by a factor $1-\frac{|\phi|^2}{v^2}$. To determine this factor, they used the condition that the Euler-Lagrange equations of the modified scalar model must coincide with those of the original one. However, since the theory is to be gauged, this is perhaps too restictive a requirement. Instead, one can consider a more general version of the non-linear Schr\"{o}dinger Lagrangian in 1+1 dimensions of the form
\beq
{\cal L}=\frac{i}{2}H(\varphi)[\phi^*\partial_0\phi-\phi\partial_0\phi^*]-\partial_x\phi^*\partial_x\phi-U(\varphi)\label{eq:24}
\eeq
where  $U$ is a potential that includes an asymmetric vacuum of modulus $v$. From (\ref{eq:24}) we obtain the conserved current
\beq
\rho=\varphi^2H(\varphi),\;\;\;\;\;\;\; j_x=-i(\phi^*\partial_x\phi-\phi\partial_x\phi^*)
\eeq
and the field momentum,
\beq
P=\frac{i}{2}\int dxH(\varphi)[\phi^*\partial_x\phi-\phi\partial_x\phi^*]
\eeq
whose variation is given by
\begin{eqnarray}
\delta P&=&i\int dx\{[H\partial_x\phi-\partial_x(H\phi)]\delta\phi^*-[H\partial_x\phi^*-\partial_x(H\phi^*)]\delta\phi\}\nonumber\\ &+&\frac{i}{2}\int dx\{\partial_x[H\phi^*\delta\phi-H\phi\delta\phi^*]+\frac{dH}{d\varphi^2}[\phi\delta\phi^*+\phi^*\delta\phi]\}
\end{eqnarray}
The difficulties emphasized in \cite{bib:bh} are twofold: if $\phi(\pm\infty)=ve^{i\chi_\pm}$ and $H(\varphi)=1$, the charge $Q=\int dx\rho$ diverges and the momentum variation includes a term $v^2[\delta\chi_+-\delta\chi_-]$  that cannot be differentiated with respect to $\delta\phi$ or $\delta\phi^*$ and therefore the canonical formalism is perturbed. However, it suffices to introduce any $H(\varphi)$  such that $H(v)=0$ to avoid both problems: $\delta P$ will then be well defined and $Q$ will be finite and, in particular, will  vanish for all vacua.

We now turn to the gauge theory and propose the following modified Jackiw-Pi model
\begin{eqnarray}
S&=&\int d^3x\{\frac{i}{2}H(\varphi)[\phi^*D_0\phi-\phi D_0\phi^*]+\frac{\kappa}{4}\varepsilon^{\alpha\beta\gamma}A_\alpha F_{\beta\gamma}-\nonumber\\ &-&\frac{1}{4}G(\varphi)F_{ij}F_{ij}-\frac{1}{2}D_k\phi^*D_k\phi-\frac{\lambda\kappa^2}{8e^2G(\varphi)}(\varphi^2-v^2)^2\}\label{eq:28}
\end{eqnarray}
where we follow the clever idea of Manton \cite{bib:mant} of taking advantage of the Galilean invariance to include an asymmetric Maxwell term without any contribution from the electric field. Nevertheless, we avoid use of external couplings to maintain the gauge invariance of the theory explicit. Furthermore, we use a dielectric function $G(\varphi)$ to build up a non-minimal interaction between the scalar and gauge fields, as is  done in \cite{bib:ln}. Below, we shall treat this issue more generally, but for the time being, we set
\beq
G(\varphi)=\frac{\kappa^2}{e^2\varphi^2}.\label{eq:patr}
\eeq

Apart from the above motives, there is another reason for including the function $H(\varphi)$ in the Lagrangian. It has recently been shown in \cite{bib:gm} that the effective theory for the low energy interaction between a planar relativistic fermionic gas and a crystalline background leads to a Chern-Simons-Higgs model in which the term in the covariant derivatives is multiplied by a function $H$; this function is fixed by self-duality and supersymmetry criteria. In a non-relativistic situation it is not necessary to use the same function for temporal and spatial derivatives and in (\ref{eq:28}) we have chosen $H=1$  for the latter. As we shall see, in a vorticial arena $|H|<1$, so that including this function as a factor only in the temporal term favours a small deformation of the vortices in their low energy motion.

Among the Euler-Lagrange equations from (\ref{eq:28}) we find the Gauss law
\beq
\kappa F_{12}-e\varphi^2H(\varphi)=0\label{eq:30}
\eeq
whose form is exactly as in (\ref{eq:3}), $\kappa F_{12}-e\rho=0$, and therefore (\ref{eq:28}) and (\ref{eq:1}) give rise to the same anyonic statistics \cite{bib:lerd}. To split (\ref{eq:28}) into kinetic and potential parts, it is convenient to adopt the temporal gauge $A_0=0$. Then,
\beq
S=\int dt(T-V)\label{eq:31a}
\eeq
\beq
T=\int d^2x\{\frac{i}{2}H(\varphi)(\phi^*\dot{\phi}-\phi\dot{\phi}^*)+\frac{\kappa}{2}\varepsilon_{kl}\dot{A}_kA_l\}\label{eq:31b}
\eeq
and $V$ coincides with (\ref{eq:4c}). Because $A_0$ is not present in (\ref{eq:31a}), the Gauss law (\ref{eq:30}) must be imposed as a constraint on the field equations arising from (\ref{eq:31a}).

We now turn to studying the static limit of the theory. Given that $V$ is the same as in the CSH model, the whole analysis of the static part of that model is still valid in our non-relativistic theory: at the self-dual limit $\lambda=\frac{e^4}{\kappa^2}$ the solutions of equations (\ref{eq:8a})-(\ref{eq:8b}) are extrema of the action (\ref{eq:28}) that, in the ${\cal C}_n$ sector, have $V=\pi v^2n$ and form the moduli space ${\cal M}_n$ with local cordinates $\{\vec{q}_a\}$ corresponding to the positions of the zeroes of $\phi$. In this system too, the self-dual solutions are absolute minima of $V$. 

The specific form of $U(\varphi)$ and $G(\varphi)$ forces the function $H(\varphi)$ to be chosen as
\beq
H(\varphi)=\frac{e^2}{2\kappa}(\varphi^2-v^2)\label{eq:32}
\eeq
in order to make the Gauss law (\ref{eq:30}) compatible with the vortex equations (12)-(13). It is remarkable that an identical choice of $H(\varphi)$ allows one to extend the generalization of the CSH system studied in \cite{bib:gm} to a N=2 SUSY theory.
\subsection{First-order vortex dynamics.}
Introduction of $H(\varphi)$ into (\ref{eq:31b}) and use of the Gauss law give
\beq
T=\int d^2x\{\frac{\kappa}{2}\varepsilon_{kl}\dot{A}_kA_l-\frac{\kappa}{2e}\dot{\Theta}F_{12}\}\label{eq:33}
\eeq
i.e., on the moduli space, $T$ and the linear term in time derivatives in (\ref{eq:4b}) agree. This guarantees the gauge invariance of (\ref{eq:33}). Additionally, if in particular we work in the gauge $\Theta (\vec{x};\vec{q}_a)=2\sum_{a=1}^n\theta (\vec{x}-\vec{q}_a)$, the Manton approach, after the algebra already seen for the CSH model, leads to
\beq
L=-\frac{2\pi\kappa}{e}\sum_{b=1}^n\dot{q}_b^kA_k(\vec{q}_b;\vec{q}_a)-V(\vec{q}_a)\label{eq:34}
\eeq
where of course $A_k(\vec{q}_b;\vec{q}_a)$ is given by (\ref{eq:14}). Because (\ref{eq:34}) is linear in the time-derivatives, the adiabatic limit is now completely satisfactory; there are no terms containing time-derivatives in the field equations that become unimportant when time goes to $\infty$ at different rates, see \cite{bib:uh} for a conceptual analysis of this situation.

In order to obtain non-trivial dynamics we must consider the ``almost'' self-dual regime \cite{bib:stua},\cite{bib:mant} i.e., we take $\lambda=\frac{e^4}{\kappa^2}+\mu$, $\mu\simeq 0$ and hence the vortices are subject to small static forces: in (\ref{eq:34}) $V(\vec{q}_a)=n\pi v^2+\mu W(\vec{q}_a)$, where
\beq
W(\vec{q}_a)=\frac{1}{8}\int d^2x\varphi^2(\vec{x};\vec{q}_a)[\varphi^2(\vec{x};\vec{q}_a)-v^2]^2
\eeq
Even though the precise expression of $W$ cannot be found analytically, it is obvious from (\ref{eq:7}) that if $\mu>0$ $(\mu<0)$  the energy of an assembly of several vortices increases (decreases) as compared with the self-dual case and this leads to the assumption that forces among vortices are repulsive (attractive) and therefore that $W(\vec{q}_a)$ is smaller for larger intervortex distances, a conjecture that is supported by numerical computations \cite{bib:sah} and theoretical arguments \cite{bib:jft}.

To appreciate the features of the dynamics derived from (\ref{eq:34}) it is convenient to analyze the $n=2$ case in some detail. We fix the center of mass of the system to be the origin of coordinates, $\vec{Q}=\frac{1}{2}(\vec{q}_1+\vec{q}_2)=0$ and work with the relative coordinate $\vec{q}=\frac{1}{2}(\vec{q}_1-\vec{q}_2)$. Notice that $\vec{b}_1=-\vec{b}_2$. To check this property one recalls that vortex indistinguishability requires that $\varphi (-\vec{q}+\vec{y};\vec{q}_a)=\varphi (\vec{q}-\vec{y};\vec{q}_a)$ and one then uses (\ref{eq:13p}). Furthemore, because the system has to be symmetric against parity and rotations around the centre of masses, 
\beq
\vec{b}_1=-\vec{b}_2=\frac{1}{2}b(q)\vec{q}\label{eq:34b}
\eeq
Introducing (\ref{eq:34b}) into (\ref{eq:34}) we find
\beq
L=-\frac{2\pi\kappa}{e^2}[\frac{1}{q^2}-b(q)]\varepsilon_{kj}\dot{q}^kq^j-\mu W(q)   .
\eeq
In terms of the polar angle $\theta=\theta(\vec{q})$
\beq
L=\frac{2\pi\kappa}{e^2}[1-q^2b(q)]\dot{\theta}-\mu W(q)
\eeq
and the dynamical equations are,
\begin{eqnarray}
\dot{q}&=&0\nonumber\\\dot{\theta}&=&-\frac{\mu e^2}{2\pi\kappa}\frac{\frac{dW}{dq}}{2qb(q)+q^2\frac{db}{dq}}  ,
\end{eqnarray}
($b(q)$ is not equal to $\frac{1}{q^2}$, see later). 
Hence, the vortices move in circular orbits with constant angular speed. The magnitude of the
angular  speed is a function of the orbit radius and the sense of movement is opposite for type I and type II superconductors.

To finish this subsection, we note that although we have fixed $G(\varphi)$ in (\ref{eq:patr}) to make the moduli space of the model fit in with that of the CSH one, a similar treatment can be equally carried out for general $G(\varphi)$. The only difference is that expression (\ref{eq:32}) must be substituted by
\beq
H(\varphi)=\frac{\kappa}{2\varphi^2}\frac{\varphi^2-v^2}{G(\varphi)}
\eeq
and that the self-duality equations are not (\ref{eq:8a})-(\ref{eq:8b}) but rather a generalization that is to be addressed in the next section. Nevertheless, (\ref{eq:33}) and the subsequent results remain valid. All the vortices of the complete family of generalized non-relativistic models (\ref{eq:28}) have exactly the same first-order dynamics.
\subsection{The effect of a charged background.}
All the generalized non-relativistic models governed by the action (\ref{eq:28}) can be modified by adding a charged constant background:
\beq
S_B=e\int d^3x v^2A_0(\vec{x}, t)
\eeq
which leads to a new Gauss law
\beq
\kappa F_{12}=-e(v^2+\varphi^2H(\varphi))\label{eq:j4}
\eeq
that renders the system self-dual at the static limit if $\lambda=\frac{e^4}{\kappa^2}$. The potential energy and Bogomolnyi equations are respectively (\ref{eq:41c}) and (\ref{eq:47a})-(\ref{eq:47b}) in this system, as we shall see in the next section. The first order equations for a general choice of $G(\varphi)$ are compatible with (\ref{eq:j4}) if and only if $H$ is chosen in the form:
\beq
H(\varphi)=\frac{v^2}{\varphi^2}[\frac{\kappa}{2G(\varphi)}(1-\frac{\varphi^2}{v^2})-1]
\eeq
The price to be paid is a different choice of $H$. For instance, the 
model discussed by Manton in Reference \cite{bib:mant} corresponds to 
${\displaystyle G(\varphi)=\frac{\kappa}{2}}$ and $H(\varphi)=-1$. 
The generalization of the system proposed in Section \S.3 obeys :
\beq
G(\varphi)=\frac{\kappa^2}{e^2\varphi^2},\ \ \ \ \ \ 
H(\varphi)=\frac{e^2}{2\kappa}(v^2-\varphi^2)-\frac{v^2}{\kappa^2}
\eeq
which it is interesting to compare with formulas (\ref{eq:patr}) and (\ref{eq:30}).

Using the Gauss law, (\ref{eq:j4}) we see that the kinetic energy 
becomes:
\beq
T=\int d^2x\{\frac{\kappa}{e}\dot{\Theta}F_{12}-\frac{\kappa}{2}\varepsilon_{kl}
\dot{A}_kA_l+v^2\dot{\Theta}\} \label{eq:j7}
\eeq
There is a new term with respect to the kinetic energy in the absence of 
background, but before analysing the physics coming from it, it is 
convenient to compare the developments of Section \S.3 with the parallel 
study in Reference \cite{bib:mant}.

If we look at our choice of gauge 
$\Theta=2\sum_{a=1}^n\theta(\vec{x}-\vec{q}_a)$ near the center of 
each vortex $\vec{x}=\vec{q}_b+\vec{\epsilon}$ and take the limit 
$|\vec{\epsilon}|\rightarrow 0$, we find:
\beq
\lim_{\epsilon\rightarrow 0}\Theta(\vec{x};\vec{\epsilon})=2\lim_{\epsilon\rightarrow 
0}\sum_{a\neq 
b}\theta(\vec{q}_b+\vec{\epsilon}-\vec{q}_a)+2\lim_{\epsilon\rightarrow 
0}\theta(\vec{q}_b+\vec{\epsilon}-\vec{q}_b)
\eeq
Solving the ambiguity by defining $\theta_b=\lim_{\epsilon\rightarrow 
0}\theta(\vec{q}_b+\vec{\epsilon}-\vec{q}_b)$, one sees that
\beq
\Theta(\vec{q}_b)=2\sum_{a\neq 
b}\theta(\vec{q}_b-\vec{q}_a)+2\theta_b=2\psi_b+2\theta_b
\eeq
which is exactly the Manton choice of gauge. To see how to glue these local choices 
it suffices to look at the case of two vortices. Near the center of the 
first vortex we have:
\beq
\Theta(\vec{x})_{\vec{x}\rightarrow\vec{q}_1}\simeq\theta(\vec{x}-\vec{q}
_1)+\psi_1,\ \ \ \ \ 
\Theta(\vec{q}_2)=\theta(\vec{q}_2-\vec{q}_1)+\psi_1
\eeq
Around the second vortex, there are similar expressions:
\beq
\Theta(\vec{x})_{\vec{x}\rightarrow\vec{q}_2}\simeq\theta(\vec{x}-\vec{q}_2)+\psi_2,\ \ \ \ \ 
\Theta(\vec{q}_1)=\theta(\vec{q}_1-\vec{q}_2)+\psi_2
\eeq
But in ${\cal M}_2$ the two descriptions above are equivalent: the 
impossibility of distinguishing the vortices requires 
$\Theta(\vec{q}_1)=\Theta(\vec{q}_2)$ and this identity determines the 
gluing by setting $\theta(\vec{q}_1-\vec{q}_2)=\psi_2$ and 
$\theta(\vec{q}_2-\vec{q}_1)=\psi_1$.

In consequence, we have found the same first-order dynamics as Manton, 
independently of the model under scrutiny. In systems that generalize the model analyzed 
in \S.3, the kinetic energy includes the first two terms of (\ref{eq:j7}), and the reduced Lagrangian is:
\begin{eqnarray}
L&=&-\frac{2\pi\kappa}{e}\sum_{b=1}^n\varepsilon_{kj}[\sum_{a\neq 
b}\dot{q}_b^k\frac{q_b^j-q_a^j}{|\vec{q}_b-\vec{q}_a|^2}-\dot{q}_b^kb_b^j
]\nonumber\\ 
&=&\frac{2\pi\kappa}{e}\sum_{b=1}^n\frac{d\psi_b}{dt}+\frac{2\pi\kappa}{e}\sum_{b=1}^n\varepsilon_{kj}b_b^j(q)\dot{q}_b^k
\end{eqnarray}
The other contribution in (\ref{eq:j7}) due to the existence of a 
constant background leads to the reduced kinetic energy;
\beq
T=\sum_{b=1}^n\int_{\Sigma}dsdt 
\hat{J}^b_{kj}[\gamma]\frac{dq_b^k}{ds}\wedge\frac{dq^j_b}{dt}=\sum_{b=1}^n\int_{\gamma=\partial\Sigma} dt \hat{a}_j^b[\gamma]\frac{dq^j_b}{dt}\label{eq:j11}
\eeq
for a motion in the $n$-vortex moduli space along a closed path $\gamma$ 
in ${\cal M}_n$. Here,
\beq
\hat{J}^b_{kj}[\gamma]=\frac{\partial\hat{a}_j^b}{\partial 
q^k_b}-\frac{\partial\hat{a}_k^b}{\partial q^j_b},\ \ \ \ \ \ 
\hat{a}_j^b(\vec{q}_b)=\varepsilon_{jk}q^b_k
\eeq
is the complex structure inherited from the field dynamics by ${\cal M}_n$ at the adiabatic limit. The contribution of (\ref{eq:j11}) is 
therefore the area $\Sigma$ enclosed by the loop $\gamma$ in ${\cal M}_n$.

Here we do not repeat Manton's derivation of this fact because there are 
no differences in the generalized models under discussion. We observe, 
however, that the action of the mechanical system is of the form
\beq
T=\sum_{b=1}^n\{\int_\gamma dt  a_j^b[\gamma]\frac{dq_b^j}{dt}+\frac{2\pi\kappa}{e}\int_\gamma dt\frac{d\psi_b}{dt}\}\label{eq:aah}
\eeq
where
\beq
a_j^b[\gamma]=\hat{a}_j^b[\gamma]+\frac{2\pi\kappa}{e}\varepsilon_{jk}b
_b^k[\gamma] .
\eeq
\section{The generalized Abelian Higgs model.}
\subsection{Self-dual vortices in the generalized AH model.}
A solvable adiabatic dynamics on the moduli space of vortices ${\cal M}_n$ also arises in the generalized Abelian Higgs model where the field dynamics is governed by the action:
\beq
S=\int d^3x\{-\frac{1}{4}G(\varphi)F_{\mu\nu}F^{\mu\nu}+\frac{1}{2}D_\mu\phi^*D^\mu\phi-U(\varphi)\}   .\label{eq:40}
\eeq
The system is relativistic, quadratic in time-derivatives of the fields and was proposed by Lee and Nam in reference \cite{bib:ln}. 

Because $G$ depends only on $\varphi$, gauge invariance is guaranteed. The model has been written in a generic form, with $G(\varphi)$ and $U(\varphi)$ unspecified; we only require that both functions be positive definite. There are several physical situations in which this kind of model is interesting, see \cite{bib:ln}. To identify the kinetic and potential parts of (\ref{eq:40}), we choose the temporal gauge and write the action in the form
\beq
S=\int dt(T-V)\label{eq:41a}
\eeq
\begin{eqnarray}
T&=&\frac{1}{2}\int d^2x\{G(\varphi)\dot{A}_k\dot{A}_k+\dot{\phi}^*\dot{\phi}\}\label{eq:41b}\\
V&=&\int d^2x\{\frac{1}{2}G(\varphi)F_{12}^2+\frac{1}{2}D_k\phi^*D_k\phi-U(\varphi)\}\label{eq:41c}
\end{eqnarray}
Observe  that the Abelian Higgs model corresponds to the choice
\beq
G(\varphi)=1\;\;\;\;\;\;\;U(\varphi)=\frac{\lambda}{8}(\varphi^2-v^2)^2\label{eq:42}
\eeq
The static energy $V$ of the CSH model however, is obtained by choosing
\beq
G(\varphi)=\frac{\kappa^2}{e^2\varphi^2}\;\;\;\;\;\;\;U(\varphi)=\frac{\lambda}{8}\varphi^2(\varphi^2-v^2)^2,\label{eq:43}
\eeq
but now the kinetic energy is different from the kinetic energy of the CSH system; as a consequence, the Gauss law derived from (\ref{eq:40}) as a constraint equation,
\beq
\partial_k[G(\varphi)F_{0k}]-e{\rm Im}(\phi D_o\phi^*)=0   ,
\eeq
also differs from the Chern-Simons Gauss law; the electric charge is not the source of the magnetic field and exotic statistics do not develop in this model. 

In any case, a configuration space ${\cal C}=\cup_{n\in{\bf Z}}{\cal C}_n$ corresponds to every $V$ of the form (\ref{eq:41c}), such that $U$ gives rise to an asymmetric vacuum. Each field configuration in ${\cal C}_n$ has quantized magnetic flux: $e\Phi_M=2\pi n$. Furthermore, given any $G(\varphi)$ there exists a potential $U$ that allows for self-duality equations, \cite{bib:ln}: one immediately sees that
\beq
U(\varphi)=\frac{\lambda\kappa^2}{8e^2G(\varphi)}(\varphi^2-v^2)^2
\eeq
produces the Bogomolny splitting
\begin{eqnarray}
V&=&\int d^2x\{\frac{1}{2}[\sqrt{G(\varphi)}F_{12}\mp\frac{e}{2\sqrt{G(\varphi)}}(\varphi^2-v^2)]^2+|D_1\phi\pm iD_2\phi|^2+\nonumber\\
 &+&\frac{1}{8}(\frac{\lambda\kappa^2}{e^2}-e^2)\frac{(\varphi^2-v^2)^2}{G(\varphi)}\}\pm\frac{ev^2}{2}\Phi_M   .
\end{eqnarray}
At the critical point $\lambda=\frac{e^4}{\kappa^2}$, the bound is saturated by the solutions of the self-duality equations
\begin{eqnarray}
eF_{12}&=&\pm\frac{e^2v^2}{2G(\varphi)}(\frac{\varphi^2}{v^2}-1)\label{eq:47a}\\D_1\phi&\pm&iD_2\phi=0\label{eq:47b}
\end{eqnarray}
that have energy $V=\pi v^2 n$ if they belong to ${\cal C}_n$. 

As in the CSH model, the Higgs field corresponding to self-dual solutions has $n$ zeroes at the points $\vec{q}_a$ in the plane, and from (\ref{eq:47b}) one sees that $\phi$ behaves  near these zeroes as in the vortex solutions of the CSH model. Away from the zeroes, the equations with the upper sign take the form
\begin{eqnarray}
\nabla^2u&=&\frac{e^2v^2}{G(u)}(e^u-1)\label{eq:48a}\\eA_k&=&-\frac{1}{2}(\partial_k\Theta+\varepsilon_{kj}\partial_ju).\label{eq:48b}
\end{eqnarray}
The vortex solutions of (\ref{eq:47a})-(\ref{eq:47b}) are the Nielsen-Olesen vortices if we take option (\ref{eq:42}) or the Jackiw-Lee-Weinberg vortices if (\ref{eq:43}) is preferred. We emphasize that the same vortex equations and identical moduli space of solutions are shared by different physical systems; the physical nature and properties of the vortices depend crucially on the model. For instance, the NO vortices are neutral in the AHS but electrically charged in the Ginzburg-Landau theory of Reference \cite{bib:mant}. By the same token, JLW vortices have electric charge in the CSH system and are neutral in the generalized AHM under discussion. Henceforth, we expect different adiabatic dynamics on the moduli space, depending on the system in question. However, we can trust the hypothesis of the isomorphism of the moduli spaces of solutions of (\ref{eq:47a})-(\ref{eq:47b}) for different $G(\varphi)$, which is supported by the insensitivity to the form of $G(\varphi)$ of the local treatment of the moduli by means of index theorem techniques, see \cite{bib:wei} or \cite{bib:jlw}. In the sequel we shall admit that the moduli space of solutions of (\ref{eq:47a})-(\ref{eq:47b}) is completely determined by the zeroes $\vec{q}_a$ of the Higgs field.
\subsection{Second-order vortex dynamics: comparison with the AH model.}
In order to study the dynamics on ${\cal M}_n$, we start by fixing the gauge, i.e, by choosing the phase $\Theta(\vec{x};\vec{q}_a)$. The choice cannot be arbitrary; because we are working in the temporal gauge, the Gauss law
\beq
\partial_k(G\dot{A}_k)+\frac{1}{2}ev^2e^u\dot{\Theta}=0\label{eq:49}
\eeq
must be maintained to ensure the invariance of (\ref{eq:41b}) under gauge transformations with parameter $\Lambda (\vec{x};\vec{q}_a)$ varying on the moduli space. This is the main difference with the
Chern-Simons theories; in this case there is no freedom to choose ${\Theta}$ in ${\bf R}^2\times{\cal M}_n$. The Gauss law and the boundary conditions at the centers of the vortices and at infinity fix the gauge completely: setting $\dot{A}_k\rightarrow \delta A_k\equiv A_k(\vec{x};\vec{q}_a+\delta\vec{q}_a)-A_k(\vec{x};\vec{q}_a)$ and using (\ref{eq:48b}) we obtain from (\ref{eq:49}) the differential equation 
\beq
\frac{dG}{du}\partial_ku[\partial_k\delta\Theta+\varepsilon_{kj}\partial_j\delta u]+G\nabla^2\delta \Theta=e^2v^2e^u\delta\Theta\label{eq:410}
\eeq
which, together with the linearization of (\ref{eq:48a})
\beq
G(u)\nabla^2\delta u+\frac{dG}{du}\nabla^2u\delta u=e^2v^2e^u\delta u\label{eq:52}
\eeq
locally determines $\Gamma (\vec{x};\vec{q}_a)$ near each point of ${\cal M}_n$ and allows one to compute $\dot{u}$ and $\dot{\Theta}$ in terms of the $\dot{\vec{q}}_a$
\begin{eqnarray}
\dot{u}(\vec{x};\vec{q}_a)&=&\frac{\partial u(\vec{x};\vec{q}_a)}{\partial q_b^k}\dot{q}_b^k\label{eq:51a}\\\dot{\Theta}(\vec{x};\vec{q}_a)&=&\frac{\partial\Theta(\vec{x};\vec{q}_a)}{\partial q_b^k}\dot{q}_b^k\label{eq:51b}
\end{eqnarray} 
in a definite way.

All the time-derivatives in (\ref{eq:41a}) can be expressed in terms of $\dot{u}$ and $\dot{\Theta}$. Because both quantities are singular at the vortex centers, it is convenient to integrate over $\tilde{{\bf R}}^2={\bf R}^2-\cup_{a}\triangle_a$ if $\triangle_a$ is an infinitesimal disk surrounding the $a^{th}$ vortex. Given that even for the case $G(\varphi)\propto \varphi^{-k},k>1$, as happens for the CSH vortices, the integrand is regular everywhere (near a $m$-vortex $G\simeq r^{-mk}$ with $r=|\vec{x}-\vec{q}_a|$ but from linearization of (\ref{eq:47a}) one has $\dot{A}_k\simeq r^{km-m+1}$, hence $G\dot{A}_k\dot{A}_k\simeq r^{mk-2m+1}$), eliminating these disks from the integration domain has a negligible effect on $T$. Now
\begin{eqnarray}
G\dot{A}_k\dot{A}_k&=&-\frac{1}{2e}G\dot{A}_k(\partial_k\dot{\Theta}+\varepsilon_{kj}\partial_j\dot{u})\\\dot{\phi}^*\dot{\phi}&=&\frac{v^2}{4}e^u(\dot{u}^2+\dot{\Theta}^2)
\end{eqnarray}
and from the first equation
\begin{eqnarray}
G\dot{A}_k\dot{A}_k&=&-\frac{1}{2e}\partial_k[G\dot{\Theta}\dot{A}_k+\varepsilon_{jk}G\dot{u}\dot{A}_j]+\nonumber\\&+&\frac{\dot{\Theta}}{2e}\partial_k(G\dot{A}_k)-\frac{\dot{u}}{2e}G\dot{F}_{12}+\frac{\dot{u}\dot{A}_j}{2e}\varepsilon_{jk}\partial_kG
\end{eqnarray}
However, using (\ref{eq:49}) and (\ref{eq:47a})
\begin{eqnarray}
\frac{\dot{\Theta}}{2e}\partial_k(G\dot{A}_k)&=&-\frac{v^2}{4}e^u\dot{\Theta}^2\\\frac{\dot{u}}{2e}G\dot{F}_{12}&=&\frac{v^2}{4}e^u\dot{u}^2-\frac{\dot{u}\dot{G}}{2e}F_{12}
\end{eqnarray}
so that the final expression for the kinetic energy is
\beq
T=\frac{1}{2}\int_{\tilde{{\bf R}}^2}d^2x\{-\frac{1}{2e}\partial_k[G\dot{\Theta}\dot{A}_k+\varepsilon_{jk}G\dot{u}\dot{A}_j]+\frac{\dot{u}}{2e}[\dot{A}_j\varepsilon_{jk}\partial_kG-\dot{G}F_{12}]\}   .\label{eq:54}
\eeq
Note that the AHM is special:  in this case $T$ reduces to a contour integral and can therefore be given in terms of data localized at the center of each vortex \cite{bib:sam}. In all other cases it is necessary to integrate over all $\tilde{{\bf R}}^2$  and this cannot be accomplished without analytical knowledge of the vorticial fields.

There is still another aspect with respect to which the AHM is special: it is the only model of the type (\ref{eq:40}) whose kinetic energy is associated with a metric on ${\cal M}_n$, which is K\"{a}hler. This can be seen following a method devised by P. Ruback , as explained in\cite{bib:sam}. To address this point, it is convenient to replace our vectorial notation by the standard complex one: $z=x^1+ix^2$, $z_a=q_a^1+iq_a^2$ and $a=A_1-iA_2$.
The kinetic energy is of the form
\beq
T=\frac{1}{2}g_{z_az_b}\dot{z}_a\dot{z}_b+g_{z_az_b^*}\dot{z}_a\dot{z}_b^*+\frac{1}{2}g_{z_a^*z_b^*}\dot{z}_a^*\dot{z}_b^*   .
\eeq
Notice that, as already mentioned, due to (85) $g$ cannot be expressed in closed form except in the AHM.
In any case ${\cal M}_n$ has a natural complex structure
\beq
\begin{array}{rccc}J:&T{\cal M}_n&\rightarrow&T{\cal M}_n\\ &\{\dot{z}_a\}&\rightarrow&\{i\dot{z}_a\}.\end{array}\label{eq:55}
\eeq
On the other hand, from (\ref{eq:48b}) and the exponential expression for $\phi$, it is easy to see that
\begin{eqnarray}
\dot{\phi}&=&\phi\eta\label{eq:50a}\\ e\dot{a}&=&i\partial_z\eta^*\label{eq:50b}
\end{eqnarray}
with $\eta=\frac{1}{2}(\dot{u}+i\dot{\Theta})$ and hence $T{\cal M}_n$ can be identified with the space of $\eta$ deformations. Although the complete determination of $\eta$ corresponding to some given $\dot{z}_a$ is not possible, we are at least able to write it as
\beq
\eta=-\sum_{a=1}^n\dot{z}_a\beta_a(z,z^*;z_a,z^*_a)\label{eq:53}
\eeq
where
\beq
\beta_a(z,z^*;z_a,z_a^*)\simeq\frac{1}{z-z_a}=\frac{1}{|\vec{x}-\vec{q}_a|}e^{-i\theta(\vec{x}-\vec{q}_a)}\label{eq:epjc}
\eeq
for $z$ very close to $z_a$. To do this, we have used only the linearity of (\ref{eq:410}) and (\ref{eq:52}) and the regularity of $\dot{\phi}$ on all ${\bf R}^2$. To prove that the coefficients in (\ref{eq:53}) are precisely $\dot{z}_a$, it is enough to solve $\phi+t\dot{\phi}=0$. Then, the complex structure (\ref{eq:55}) is equivalent to
\beq
J\eta=i\eta.
\eeq
Now, from (\ref{eq:41b}) the metric on ${\cal M}_n$ can be recast as
\beq
g(\eta_1,\eta_2)=\frac{1}{4}\int d^2x\{G(\varphi)[\dot{a}^*_1\dot{a}_2+\dot{a}^*_2\dot{a}_1]+\dot{\phi}^*_1\dot{\phi}_2+\dot{\phi}^*_2\dot{\phi}_1\}
\eeq
where both $\dot{a}_r$ and $\dot{\phi}_r$ come from $\eta_r$ by using (\ref{eq:50a})-(\ref{eq:50b}). $g$ is clearly hermitian, $g(J\eta_1,J\eta_2)=g(\eta_1,\eta_2)$ and its K\"{a}hler form $\omega(\eta_1,\eta_2)=g(J\eta_1,\eta_2)$ is 
\beq
\omega=\frac{i}{4}\int d^2x\{G(\varphi)da^*\wedge da-d\phi^*\wedge d\phi\}   .
\eeq
It is easy to compute the exterior derivative of $\omega$
\beq
d\omega=\frac{i}{4}\int d^2x\{\frac{1}{2}\varphi\frac{dG}{d\varphi}[d\eta+d\eta^*]\wedge da^*\wedge da\}
\eeq
because $d\phi=\phi d\eta$ is an element in $\Lambda{\cal M}_n$. Therefore, $\omega$ is closed only if $G$ is constant, i.e. for the AHM.
\subsection{Vortex scattering: comparison with the CSH model.}
From formula (\ref{eq:54}) we have seen the difficulty involved in finding an exact closed expression for $T$ when $G\neq {\rm constant}$. If the vortices are close enough it is possible, 
however, to obtain a picture of the scattering that is essentially 
correct. In the case of $n=2$, the space of the polynomials 
$P_2(z)=(z-z_1)(z-z_2)=z^2+a_1z+a_2$ is isomorphic to ${\cal M}_2$. 
Notice that $z_1,z_2$ are the vortex centres and ${\cal M}_2$ is the set 
of unordered pairs of points in the plane: given $(a_1,a_2)$, we have 
either $(z_1=z_+,z_2=z_-)$ or $(z_1=z_-,z_2=z_+)$, where 
$z_\pm=\frac{(a_1\pm \sqrt{a_1^2-4a_2})}{2}$. In the center of mass 
system, $(a_1=0,a_2=w)$ implies
$P_2^R(z)=(z-\sqrt{w})(z+\sqrt{w})$. The motion is symmetric around 
the $CM$ and, when $w\rightarrow 0$, the two vortices tend to overlap at the origin. Reciprocally, we can use (\ref{eq:50a}) to express the scalar field of a system of two neighbouring vortices as
\beq
\phi(z,z^*;t)=\phi^{(2)}(z,z^*)-w(t)\phi^{(2)}(z,z^*)\beta(z,z^*)   
.\label{eq:58}
\eeq
where, $w,\dot{w}$ are small, $\phi^{(2)}$ is the radial 2-vortex 
solution, and $\beta$, accounting for the splitting of the two vortices 
, behaves as $\beta\simeq\frac{1}{z^2}$ near $z\simeq 0$, see (\ref{eq:epjc}). Hence, we see that $\phi (z,z^*;t)=0$ has the symmetric roots 
$z_1(t)=\sqrt{w(t)}$, $z_2(t)=-\sqrt{w(t)}$ around the origin, fitting with the above description in terms of $P^R_2(z)$. From (\ref{eq:58})
\beq
\dot{\phi}=-\dot{w}\phi^{(2)}\beta
\eeq
and by comparison with (\ref{eq:50a}),(\ref{eq:50b}) we know that
\beq
e\dot{a}=i\dot{w}\partial_z\beta^*
\eeq
and hence the kinetic energy is
\beq
T=\frac{1}{2}|\dot{w}|^2\int 
d^2x\{\frac{G}{e}|\partial_z\beta^*|^2+|\phi^{(2)}\beta|^2\}\equiv\frac{1
}{2}M|\dot{w}|^2\label{eq:59}
\eeq
while $M$ is given in terms of the fields of the radial 2-vortex and the 
deformation $\beta$ coming from (\ref{eq:410}) and (\ref{eq:52}). The 
form of (\ref{eq:59}) as a function of the relative coordinate 
$z_r(t)=\sqrt{w(t)}$ is
\beq
T=2M|z_r|^2|\dot{z}_r|^2.\label{eq:103}
\eeq

However, to study the movement of non-distant vortices it is more 
convenient to use (\ref{eq:59}) directly. Because (\ref{eq:59}) is the kinetic energy of a free particle in the $w$-plane, the radial 
trajectories crossing the origin are solutions of the dynamics. Nevertheless, in view of 
the equation $z_r(t)=\sqrt{w(t)}$ we find the correspondence shown in 
 Figure.1, and the celebrated $90^\circ$ scattering appears as a 
generic feature of the models (\ref{eq:40}). Note that written in the 
good coordinate in the moduli space, the $w$, the metric is flat near 
the point where the two vortices overlap, showing that the manifold ${\cal 
M}_2$ is smooth at this point and that the conical singularity suggested by 
(\ref{eq:103}) is only an artifact of having chosen the wrong relative coordinate. In 
fact, we do not expect that the abrupt change in direction shown in 
Fig.1 actually occurs, the reason being that (\ref{eq:59}) is an 
asymptotic expression valid only for very small intervortex distances. 
In a realistic scattering, the initial separation between the two 
vortices is enough to bring the subdominant contributions not included in 
(\ref{eq:59}) into play. These in turn give rise to interactions that 
produce the smooth bending of the trajectory and the situation depicted 
in Fig.1 is only reached asymptotically.

It is interesting at this point to study the vortex motion under these 
conditions in the CSH system where the term
\beq
L^{(1)}=-\frac{2\pi\kappa}{e^2}[\frac{1}{q^2}-b(q)]\epsilon_{kj}\dot{q}
^kq^j
\eeq
leads to a slightly modified kinetic energy \cite{bib:dzi2}:
\beq
T=\frac{m}{2}|z_r|^2|\dot{z}_r|^2+\frac{2\pi\kappa}{e^2}c|z_r|^4\dot{\theta}_r
\eeq
where $\theta_r=\arg z_r$, $m=2\int dx^2|\phi^{(2)}\beta|^2$ and 
$b(q)\simeq \frac{1}{q^2}+cq^2$.
The expansion of the deformation factor $b(q)$ induced by the 
interaction with the other vortices around the point $q=0$ differs in 
the CSH system from that of the AHM. An indirect argument suggests that 
the tedious computation leading to such a result is correct. Unlike in 
the case of the Ginzburg-Landau vortices, $b(q)-\frac{1}{q^2}$ cannot be 
constant at $q=0$ in the CSH model due to the nature of the Higgs 
potential ruling the interactions.
 Again, the $w$-coordinate is better suited to describing the vortex motion, 
and we find
\beq
T=\frac{\mu}{2}(|\dot{w}|^2+|w|^2\dot{\chi}^2)+\gamma|w|^2\dot{\chi}
\eeq
where $\mu=\frac{m}{8}$, $\gamma=\frac{\pi\kappa c}{e^2}$ and 
$w=|w|e^{i\chi}$. There is also a term causing ninety degree 
scattering. The new linear term in $\dot{\chi}$, however, 
completely modifies this behaviour. An intrinsic angular momentum is induced by 
this term:
\beq
J=\frac{\partial T}{\partial \dot{\chi}}=\mu|\omega|^2\dot{\chi}+\gamma|\omega|^2
\eeq
which is conserved; $\dot{J}=0$ because $\frac{\partial T}{\partial \chi}=0$. The energy of this mechanical system is:
\beq
H=\frac{1}{2}\mu|\dot{\omega}|^2+\frac{(J^2-\gamma|\omega|^2)^2}{2\mu|\omega|^2}
\eeq
which is equivalent to an isotropic harmonic oscillator. Choosing the 
constant of motion as $J=j$, we
have
\beq
H=\frac{1}{2}\mu|\dot{\omega}|^2+\frac{j^2}{2\mu|\omega|^2}+\frac{\gamma^2|\omega|^2}{2\mu}-\frac{j\gamma}{\mu}    .
\eeq
All the trajectories are thus ellipses and the motion corresponds to 
bound states of two-vortices orbiting around each other. This is 
consistent with what was discussed in section \S 2 to the effect that the inertia 
of a CSH-vortex is smaller than its mass: the vortices are trapped 
forming bound states as a result of the first-order dynamics.

In fact, modifications due to higher order terms in the expansion of 
$b(q)$, to be taken into account at larger intervortex distances, do not 
alter this picture. The energy and angular momentum would in this 
case be:
\beq
H=\frac{1}{2}\mu|\dot{\omega}|^2+\frac{(J^2-h(|\omega|)^2}{2\mu|\omega|^2}
\eeq
\beq
J=\mu|\omega|^2\dot{\chi}+h(|\omega|)
\eeq
where $h(|\omega|)$ is a power series in $|\omega|^2$. From 
$\dot{H}=\dot{J}=0$ one reads the motion equations:
\beq
\mu|\ddot{\omega}|-\frac{1}{\mu|\omega|^3}(J-h(|\omega|)(J-h(|\omega|)+|\omega|h^{\prime}(|\omega|))=0
\eeq
\beq
\mu|\omega|\ddot{\chi}+2\mu|\dot{\omega}|\dot{\chi}+\frac{h^{\prime}(|\omega|)|\dot{\omega}|}{|\omega|}=0    .
\eeq
Circular trajectories , $|\omega|=a$, occur if 
$j=h(a)-ah^{\prime}(a) $ with angular velocity: 
$\dot{\chi}=-\frac{h^{\prime}(a)}{2a\mu}$. Vortex bound states do not 
arise only at short distances.
\section{Conclusions and outlook.}
Implementation of the Manton approach to the low-speed dynamics of 
the topological vortices in the CSH model is too involved to allow a 
successful analytical treatment. Nevertheless, we have shown that it is 
possible to build two different kinds of self-dual generalized Abelian 
Higgs systems with solvable slow vorticial dynamics and whose parameters 
can eventually be adjusted to obtain exactly the CSH moduli space. 
Remarkably enough, despite important differences in their field 
profiles, the qualitative dynamical behavior of the vortices in each 
class of generalized systems is not particularly model-dependent but is 
generic: all the possible non-relativistic first order systems give rise 
to a uniform circular motion of the vortices around the barycenter and for 
all the relativistic second-order ones the head-on collision of two 
defects leads to right angle scattering. It is believed that the 
dynamics of the original CSH vortices results from some entanglement of 
these two effects.

A few final words on quantization. For the quadratic model of Section 4. the transition from classical to quantum mechanics is straightforward: the Laplace-Beltrami operator corresponding to the metric on the moduli space becomes the quantum Hamiltonian replacing the classical kinetic energy as generator of the
dynamics. In the linear model of Section 3 things are more interesting (less standard) , especially when the charged background is incorporated. Observe that (\ref{eq:aah}) is no more than topological classical mechanics associated with the 
space of paths in ${\cal M}_n$, see Jackiw et. al. \cite{bib:djt}. The 
quantization is almost trivial when ${\cal M}_n$ is topologically trivial. The Hilbert space reduces to the ground state, which is 
degenerated; e.g., if $b_b^k[\gamma]=0$, it would be the first Landau 
level. If vortices move in a compact space, a two-sphere for instance,
things become more difficult, and one would need to consider the Floer 
homology of the symplectic compact manifold ${\cal M}_n$ \cite{bib:jm}.

\newpage
\begin{figure}[h]
\begin{center}
\epsfig{file=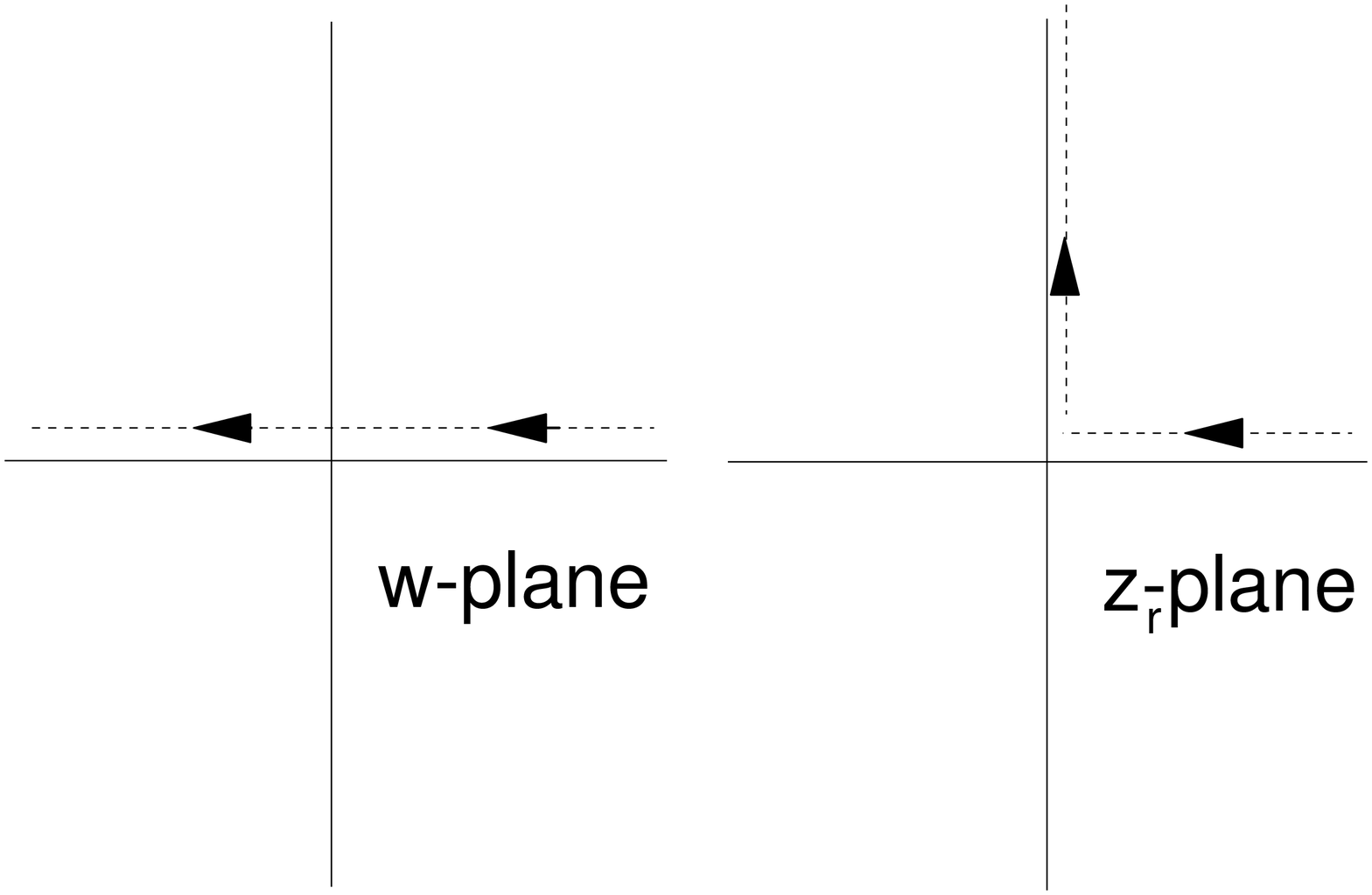,height=10cm,width=13cm}
\caption{Scattering of a system of two vortices as seen from the
squared relative coordinate and from the true relative coordinate planes.}
\end{center}
\end{figure}
\end{document}